\documentclass{article}
\usepackage{spconf,amsmath,graphicx}
\usepackage{multirow}
\usepackage{bbding}
\usepackage{pifont}
\usepackage{wasysym}
\usepackage{amssymb}
\usepackage{hyperref}
\usepackage{subfigure}

\title{Coarse-to-Fine Covid-19 Segmentation via Vision-Language Alignment}
\name{Dandan Shan$^{\star}$\qquad Zihan Li\thanks{The first two authors have the equal contribution.}$^{\dagger}$ 
\qquad Wentao Chen$^{\ddag}$ 
\qquad Qingde Li$^{\ast}$ 
\qquad Jie Tian$^{\diamond}$ 
\qquad Qingqi Hong\thanks{Corresponding author: Qingqi Hong.}$^{\star\dagger\dagger}$}
  
\address{$^{\star}$Xiamen University\quad
$^{\dagger}$UIUC\quad
$^{\ddag}$BUPT\quad
$^{\ast}$University of Hull\quad
$^{\diamond}$CAS\quad
$^{\dagger\dagger}$Hong Kong COCHE
}
\begin{document}
\topmargin=0mm
%
\maketitle
\begin{abstract}
Segmentation of COVID-19 lesions can assist physicians in better diagnosis and treatment of COVID-19. However, there are few relevant studies due to the lack of detailed information and high-quality annotation in the COVID-19 dataset. To solve the above problem, we propose C2FVL, a Coarse-to-Fine segmentation framework via Vision-Language alignment to merge text information containing the number of lesions and specific locations of image information. The introduction of text information allows the network to achieve better prediction results on challenging datasets. We conduct extensive experiments on two COVID-19 datasets including chest X-ray and CT, and the results demonstrate that our proposed method outperforms other state-of-the-art segmentation methods.
\end{abstract}
\section{Introduction}
\label{sec:intro}

Deep learning technique has been widely used in the area of medical image processing, and the segmentation of lesions using neural networks can significantly reduce time and labor costs. Since multimodal images exhibit more relevant information \cite{naik2020denouements} than unimodal images, some studies \cite{dolz2018hyperdense,fu2021multimodal} merge images of different modalities to segment the target. 
However, medical images lack detailed information, making it challenging to segment precisely only by medical images. Some works \cite{radford2021learning,lu2019vilbert,li2022lvit} use the method of image information fusion with text information for image segmentation. Although Tomar et al. \cite{tomar2022tganet} applied image and text information fusion to medical image segmentation, it did not consider the location information of target. The segmentation of the lesion area is an effective means to diagnose and treat COVID-19. However, only experienced radiologists can accurately label lesions, and with few publicly available datasets \cite{kugunavar2021convolutional}, make it difficult to improve the accuracy of segmentation networks \cite{xu2022mrdff,saeedizadeh2021covid,alom2020covid_mtnet}. To solve the above problems, we introduce text information in the process of COVID-19 segmentation. Since the text contains the number and specific location information of lesions, it can assist the network in learning more features with rich semantic information from coarse to fine on a limited and poorly labeled dataset. In summary, the main contributions of this paper are:
\begin{itemize}
\item We construct the C2FVL segmentation framework using CNN and Vision Transformer to achieve accurate segmentation of COVID-19 effectively.
\item We propose a Vision Language Alignment Module (VLAB) and a novel loss function to facilitate the alignment of text and image information, which improves segmentation accuracy.
\item We compare our C2FVL with the state-of-the-art segmentation methods on two COVID-19 datasets, and the experiments show that the performance of C2FVL is optimal. The code of our proposed C2FVL is made available at GitHub\footnote{\href{https://github.com/HUANGLIZI/C2FVL}{https://github.com/HUANGLIZI/C2FVL}}.
\end{itemize}
\vspace{-3mm}
\section{Related works}
\vspace{-2mm}
\label{sec:relatedworks}
\textbf{Multi-modal Information Fusion:} 
Multiple modality can provide complementary information, and their fusion can effectively improve the accuracy of the segmentation of medical images. Dolz et al. \cite{dolz2018hyperdense} used dense connections between different modalities to swap information.
CLIP \cite{radford2021learning} used a contrast learning approach to learn image and text information, achieving zero-shot transfer learning.  
TGANet \cite{tomar2022tganet} introduced text about the size and number of polyps, enabling automatic segmentation of polyps of different scales.
\setlength{\belowcaptionskip}{0pt}
\setlength{\abovecaptionskip}{0pt}
\begin{figure*}[!ht]
    \centering
	\centerline{\includegraphics[width=0.9\textwidth]{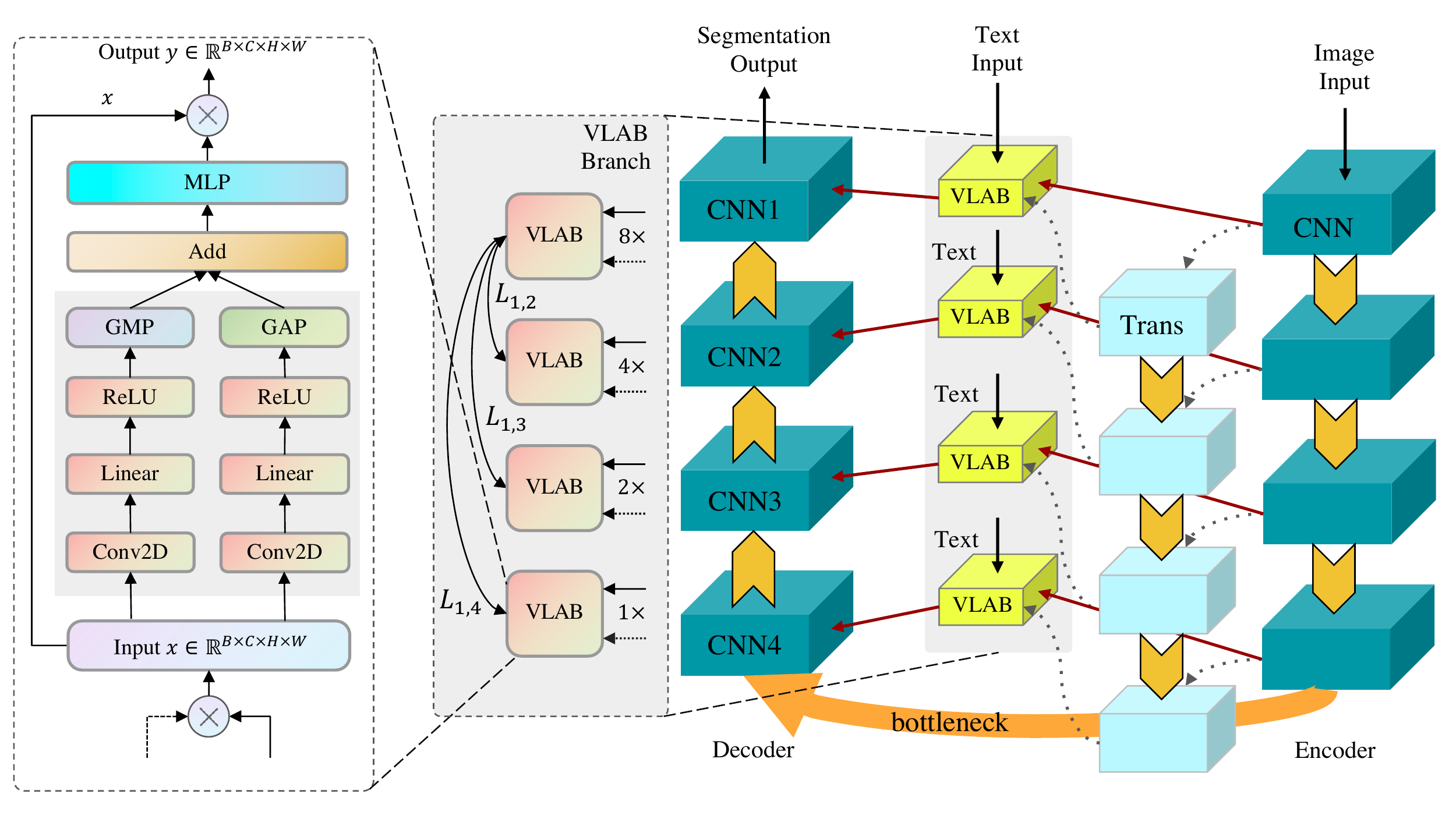}}
	\caption{Overview of the C2FVL framework. It consists of three parts, encoder, VLAB branch and decoder.}
	\label{fig1}
	\vspace{-2mm}
\end{figure*}

\noindent\textbf{COVID-19 Segmentation:} 
Saeedizadeh et al. \cite{saeedizadeh2021covid} added a regularization term to the loss function of the UNet to detect chest regions infected with COVID-19 in CT images. 
Alom et al. \cite{alom2020covid_mtnet} used the NABLA-N model to segment the COVID-19-infected region in the designed system. 

\vspace{-3mm}
\section{Method}
\label{sec:method}
\vspace{-2mm}
\subsection{Structure of C2FVL Model}
\vspace{-1mm}
As shown in Fig. \ref{fig1}, our model follows an encoder-decoder structure, where the encoder uses hierarchical CNN and Vision Transformer to extract coarse image features. At the skip connection, the text information is fused with image information and design the VLAB module to facilitate cross-modal alignment between text and images. The decoder consists of CNN that processes low-level detail information and aligned text-image features to reconstruct fine segmentation masks.

\vspace{-4mm}
\subsection{Encoder Module}
\vspace{-1mm}
For medical image segmentation tasks, both global and local features are essential. Traditional convolutional neural networks use a fixed-size receptive field to learn to local features of an image. \cite{huang2021gloria} uses average pooling to extract global features, but its capability is limited. In contrast, the Multi-headed Self-attention mechanism in the Vision Transformer works directly on the whole image and can learn more global features. Therefore, we use several CNN and Vision Transformer blocks in the encoder module to extract image features jointly. First, we take the image as the input of the CNN block and then send the feature map to the Vision Transformer after convolution operation, batch normalization, and activation function. Meanwhile, max-pooling is used to downsample the feature map to retain more texture features.
For the activation function, we use ReLU to enhance the expression ability of the network. In the Vision Transformer part,  the features extracted by the CNN are encoded to capture global features. At each stage, the scale of the output of Vision Transformer is adjusted to the same as the output of the CNN block, and adding them together to fuse local and global features, as shown in the following equations.
\vspace{-2mm}
\begin{eqnarray}
\setlength{\abovedisplayskip}{0pt}
\setlength{\belowdisplayskip}{0pt}
& F_{rt,i}= \sigma\left ( BatchNorm\left ( Conv\left ( Upsample\left ( F_{vit,i}\right )\right ) \right)\right )\label{3.2-1}\\
& F_{encoder,i}= F_{cnn,i}+F_{rt,i}\label{3.2-2}
\setlength{\belowdisplayskip}{3pt}
\end{eqnarray}
where $F_{vit,i}$ denotes the output of the $i$-th layer Vision Transformer, $F_{rt,i}$ denotes the result of the reconstruction of $F_{vit,i}$, and $F_{cnn,i}$ denotes the output of the $i$-th layer CNN, $\sigma$ is ReLU activation function.
 
\begin{figure*}[!ht]
	\centering
	\subfigure[Input]{
		\begin{minipage}[t]{0.135\linewidth}
			\centering
			\includegraphics[width=0.9in]{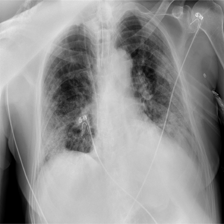}\\
			\vspace{0.02cm}
			\includegraphics[width=0.9in]{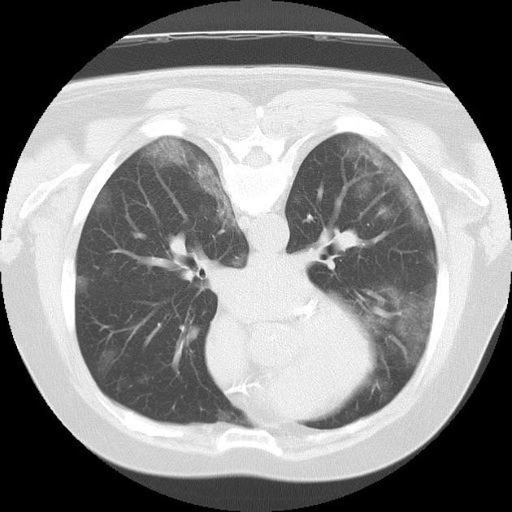}\\
			\vspace{0.02cm}
		\end{minipage}%
	}%
	\subfigure[Ground Truth]{
		\begin{minipage}[t]{0.135\linewidth}
			\centering
			\includegraphics[width=0.9in]{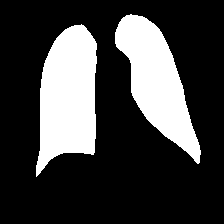}\\
			\vspace{0.02cm}
			\includegraphics[width=0.9in]{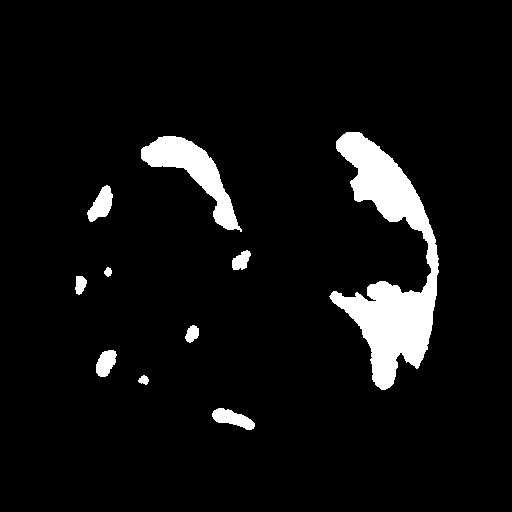}\\
			\vspace{0.02cm}
		\end{minipage}%
	}%
	\subfigure[UNet++]{
		\begin{minipage}[t]{0.135\linewidth}
			\centering
			\includegraphics[width=0.9in]{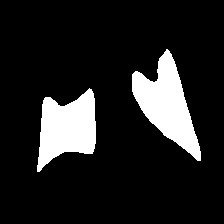}\\
			\vspace{0.02cm}
			\includegraphics[width=0.9in]{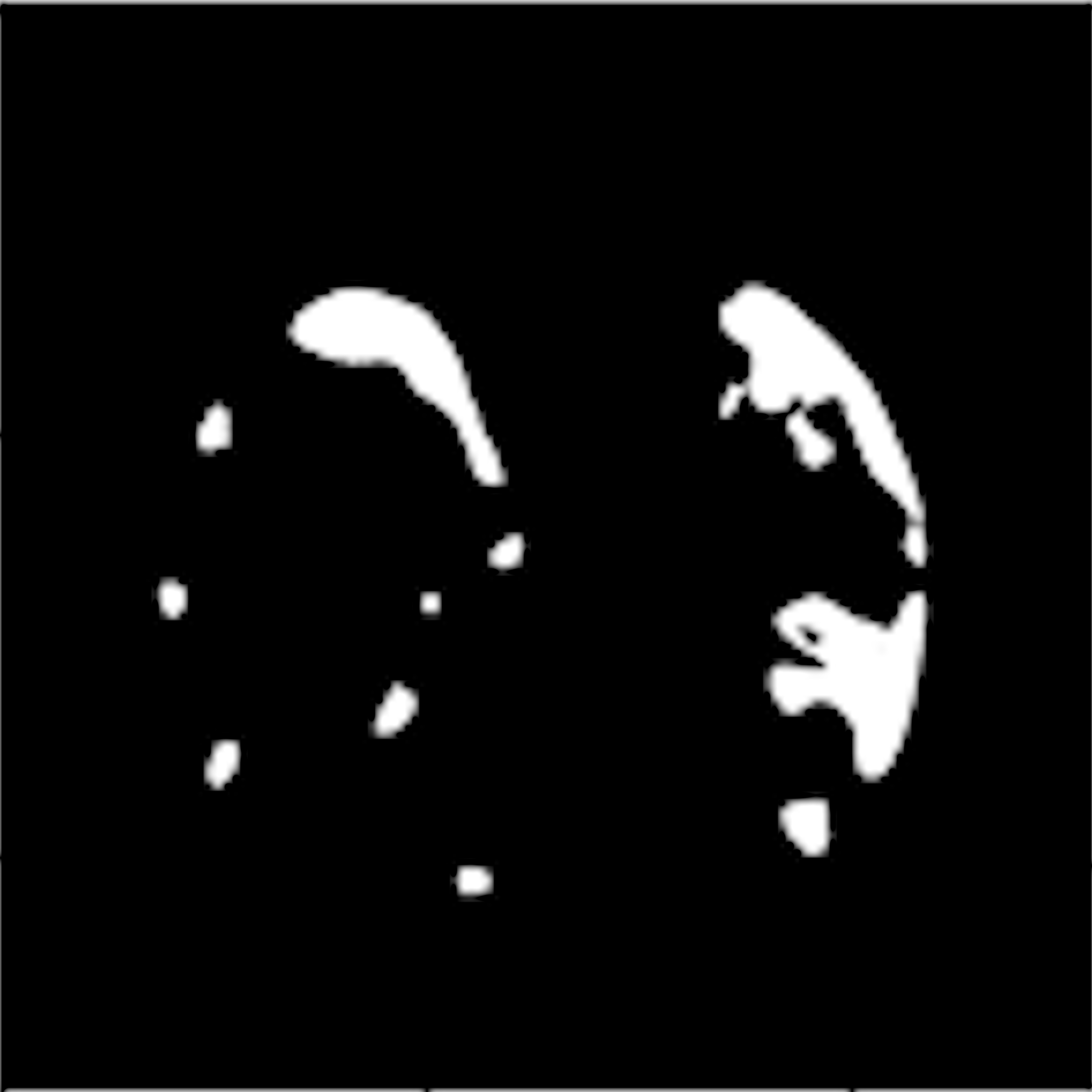}\\
			\vspace{0.02cm}
		\end{minipage}%
	}%
	\subfigure[nnUNet]{
		\begin{minipage}[t]{0.135\linewidth}
			\centering
			\includegraphics[width=0.9in]{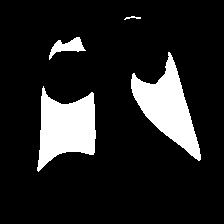}\\
			\vspace{0.02cm}
			\includegraphics[width=0.9in]{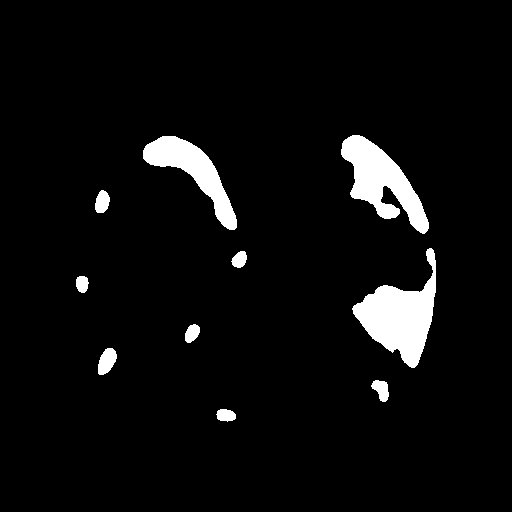}\\
			\vspace{0.02cm}
		\end{minipage}%
	}%
	\subfigure[TGANet]{
		\begin{minipage}[t]{0.135\linewidth}
			\centering
			\includegraphics[width=0.9in]{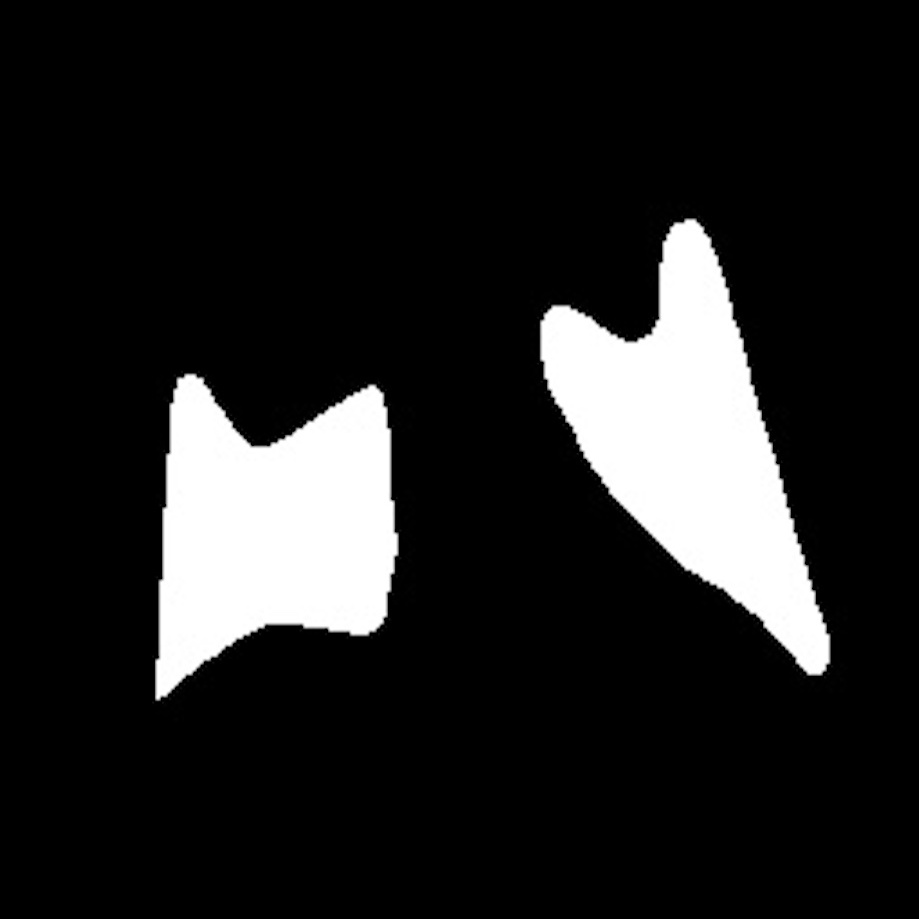}\\
			\vspace{0.02cm}
			\includegraphics[width=0.9in]{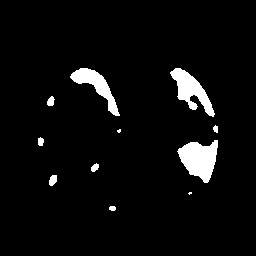}\\
			\vspace{0.02cm}
		\end{minipage}%
	}
	\subfigure[GLoRIA]{
		\begin{minipage}[t]{0.135\linewidth}
			\centering
			\includegraphics[width=0.9in]{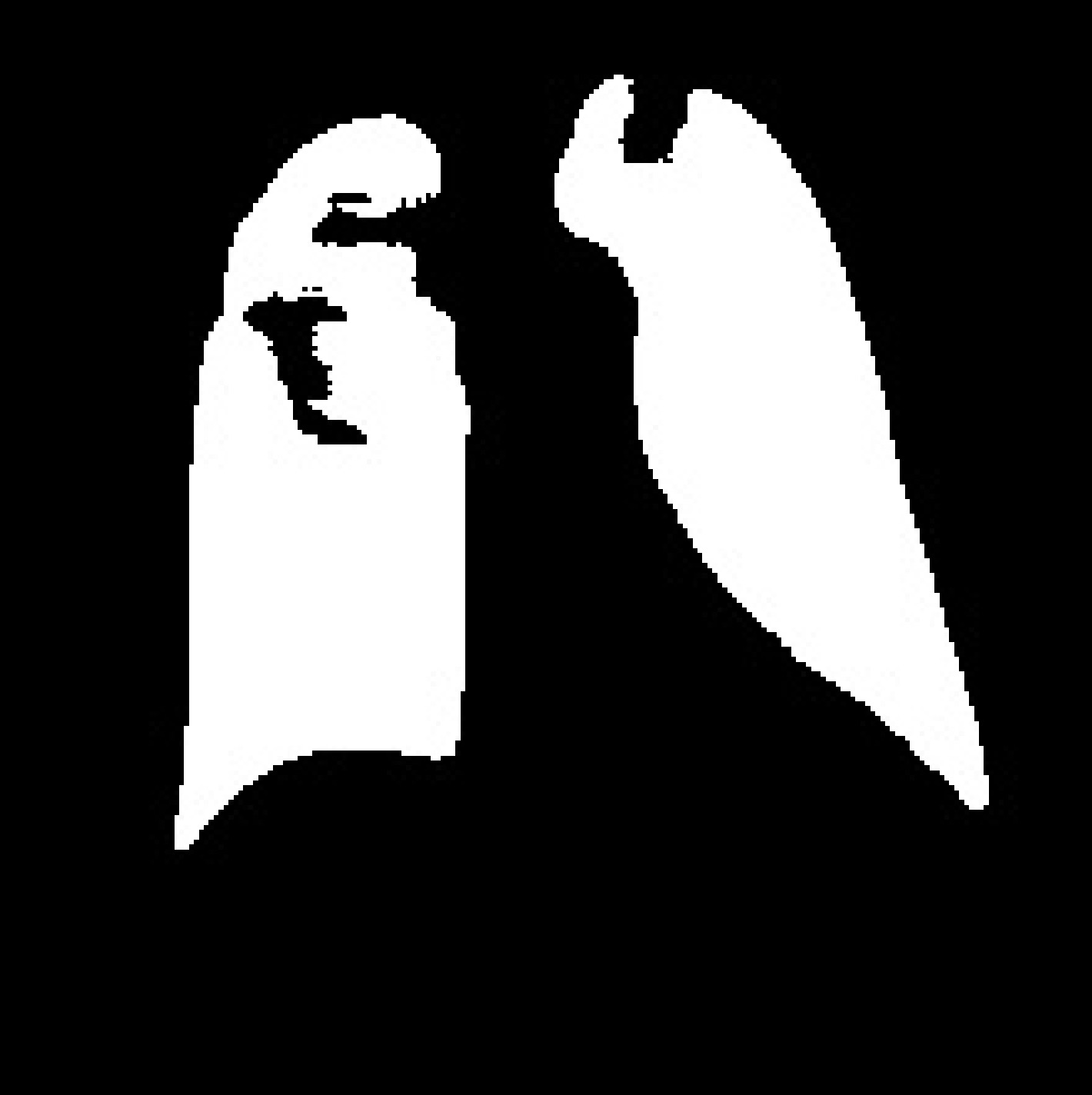}\\
			\vspace{0.02cm}
			\includegraphics[width=0.9in]{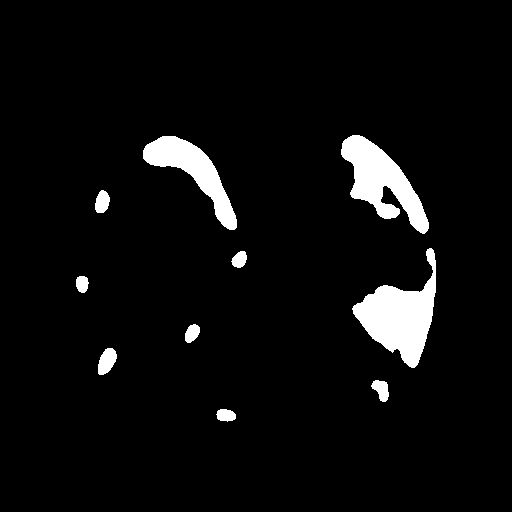}\\
			\vspace{0.02cm}
		\end{minipage}%
	}
	\subfigure[Ours (C2FVL)]{
		\begin{minipage}[t]{0.135\linewidth}
			\centering
			\includegraphics[width=0.9in]{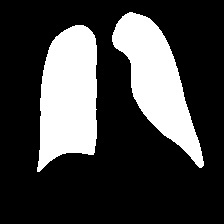}\\
			\vspace{0.02cm}
			\includegraphics[width=0.9in]{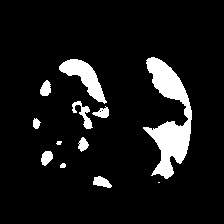}\\
			\vspace{0.02cm}
		\end{minipage}%
	}%
	\centering
	\caption{Visualization results on the QaTa-COVID dataset (row 1) and MosMedData+ dataset (row 2). From left to right: (a) input image, (b) ground-truth, (c) UNet++ and (d) nnUNet are predictions of baseline without text information, while (e) TGANet and (f) GLoRIA are predictions with text information. And (g) is the prediction of our proposed C2FVL.}
	\label{visfig}
	\vspace{-0.4cm}
	\label{vis}
\end{figure*}
\vspace{-4mm}
\subsection{Multi-sclale Vision-Language Aggregation}
\vspace{-2mm}
We convert the text information into a vector with dimension of 8. The first dimension indicates whether the lesion area is bilateral, the second dimension indicates the number of lesions, and each subsequent three dimensions indicates the location of the lesions in the left and right lungs in that order. For example, suppose the text message is ``Bilateral pulmonary infection, two infected areas, upper middle left lung and middle lower right lung''. In that case, we transform it into a vector $V_{text}=\left [1,2,1,1,0,0,1,1\right ]$. Unlike \cite{tomar2022tganet}, which uses text-guided image classification to assist in the segmentation task, we use text information to select features. Specifically, for feature maps of different scales, we employ the repeat operation to align $V_{text}$ with the number of channels of the image features and then weight the image features with text to extract the lesion's location and quantity features while suppressing other irrelevant features, as shown in the Eqn. (\ref{3.3-8}).
\vspace{-2mm}
\begin{eqnarray}
\setlength{\abovedisplayskip}{0pt}
\setlength{\belowdisplayskip}{0pt}
& F_{vl,i}= F_{encoder,i}\odot \left ( V_{text}.repeat\left (channel_{i}/8,1\right )\right )\label{3.3-8}
\end{eqnarray}

\vspace{-5mm}
\subsection{Vision-Language Alignment Block (VLAB)}
\vspace{-2mm}
Inspired by the convolutional block attention module \cite{li2022tfcns} and CUM \cite{li2022semi}, we design our Vision-Language Alignment Block (VLAB) with a parallel structure to facilitate the alignment of text and image features. As shown in Fig. \ref{fig1}, in the left and right branches, the input features are first processed using the MLP structure and then the feature maps are subjected to global max pooling (GMP) and global average pooling (GAP), respectively. The outputs of two branches are added to further enhance the same features, and finally multiplied with the input after two layers of MLP structure. The process can be defined as follows.
\vspace{-2mm}
\begin{eqnarray}
\setlength{\abovedisplayskip}{0pt}
\setlength{\belowdisplayskip}{0pt}
& F_{avg}=GlobalAvgPool\left ( MLP\left (   x\right  )\right )\\
& F_{max}=GlobalMaxPool\left ( MLP\left (   x\right  )\right )\\
& y= MLP\left (F_{avg}+F_{max}\right )\otimes  x
\end{eqnarray}
In addition, We also calculate the cosine loss between different VLAB outputs to force each network layer to focus on the same focal region and append this part of the loss after the dice loss $L_{Dice}$ and the cross entropy loss $L_{CE}$, with the designed loss function defined as shown in Eqn. (\ref{3.3-12}) and (\ref{3.3-13}).
\vspace{-2mm}
\begin{eqnarray}
\setlength{\abovedisplayskip}{0pt}
\setlength{\belowdisplayskip}{0pt}
& L_{4,i}=1-\frac{y_{4}\cdot Downsample\left ( y_{i}\right ) }{\left | y_{4}\right |\times\left | Downsample\left ( y_{i}\right )\right | }\label{3.3-12}\\
& L_{V1}=\frac{1}{2}L_{Dice}+\frac{1}{2}L_{CE}+\alpha L_{4,1}+\beta L_{4,2}+\gamma L_{4,3}\label{3.3-13}
\end{eqnarray}
where $L_{4,i}$ ($i\in 1,2,3$) denotes the cosine loss between the 4-th layer VLAB output $y_{4}$ and the $i$-th layer VLAB output $y_{i}$. The default values of $\alpha$, $\beta$, $\gamma$ are all 0.5. In addition, we also design $L_{1,i}$ ($i\in 2,3,4$) to calculate the cosine loss between the outputs of the first layer VLAB and the output of the lower layer VLAB, as shown in Eqn. (\ref{3.3-14}) and (\ref{3.3-15}).
\vspace{-2mm}
\begin{eqnarray}
\setlength{\abovedisplayskip}{0pt}
\setlength{\belowdisplayskip}{0pt}
& L_{1,i}=1-\frac{y_{1}\cdot Upsample\left ( y_{i}\right ) }{\left | y_{1}\right |\times\left | Upsample\left ( y_{i}\right )\right | }\label{3.3-14}\\
& L_{V2}=\frac{1}{2}L_{Dice}+\frac{1}{2}L_{CE}+\alpha L_{1,4}+\beta L_{1,3}+\gamma L_{1,2}\label{3.3-15}
\end{eqnarray}

\vspace{-3mm}
\section{Experiments}
\label{sec:experiments}
\vspace{-3mm}
\subsection{Setup}
\vspace{-2mm}
\noindent\textbf{Datasets:} We use open-acess QaTa-COVID dataset \cite{yamac2021convolutional} and MosMedData+ dataset \cite{morozov2020mosmeddata,COVID-19-CT} in the experiments to evaluate the performance. The training and test sets of QaTa-COVID contain 7145 and 2113 chest X-ray images with annotations, respectively. We use a 4:1 ratio to split the initial training set into training and validation. MosMedData+ contains 2729 CT scan slices of lung infections. To balance the different morphologies of the lesion, we consider the images with four prefixes as four classes. For each class, we distribute them in a ratio of 8:1:1 in the training set, validation set, and test set.

\noindent\textbf{Implementation Details:} Our experiments are conducted on four NVIDIA GeForce GTX 1080 Ti GPUs with the initial learning rate of model training set to 1e-3 and the batch size to 4. The total number of iterations is set to 2000 rounds, and the number of early stop rounds is set to 50.
\vspace{-5mm}
\begin{table}[!ht]
\centering
\caption{Performance comparison between our method (C2FVL) and other state-of-the-art methods on datasets.}
\resizebox{\linewidth}{!}{%
\begin{tabular}{cccccc}
\hline
\multirow{2}{*}{\textbf{Method}} & \multirow{2}{*}{\textbf{Text}} &\multicolumn{2}{c}{\textbf{QaTa-COVID}} & \multicolumn{2}{c}{\textbf{MosMedData+}} \\ \cline{3-4} \cline{5-6}
                        && Dice (\%)     & IoU (\%)     & Dice (\%)      & IoU (\%)     \\ \hline
UNet \cite{ronneberger2015u}                   & $\times$ &  79.02 & 69.46   & 64.60 &	50.73\\
UNet++ \cite{zhou2018unet++}                  & $\times$ & 79.62 & 70.25   & 71.75 &	58.39\\
AttUNet \cite{oktay2018attention}                 & $\times$ & 79.31 & 70.04 & 66.34 &	52.82 \\
nnUNet \cite{isensee2021nnu}                  & $\times$ & 79.89 & 70.58 & 72.59 &	60.36\\
TransUNet \cite{chen2021transunet}               & $\times$ & 78.63 & 69.13 & 71.24 & 58.44\\
Swin-UNet \cite{cao2021swin}                & $\times$ & 77.27 & 67.96 & 63.29 & 50.19
\\
UCTransNet \cite{wang2022uctransnet}               & $\times$ & 79.15 & 69.60 & 65.90 &	52.69 \\
TGANet \cite{tomar2022tganet}                  & \checkmark & 79.87 & 70.75 & 71.81 & 	59.28\\
GLoRIA\cite{huang2021gloria} & \checkmark&79.94&70.68 & 72.42& 60.18
\\\textbf{Ours} & \checkmark &\textbf{83.40}& \textbf{74.62} &\textbf{74.56}&  \textbf{61.15}\\
\hline
\end{tabular}
}
\label{tab1}
\vspace{-6mm}
\end{table}

\vspace{-2mm}
\subsection{Comparison with SOTA Methods}
\vspace{-2mm}
We conducted comparison experiments on the two datasets with state-of-the-art methods, including methods with only image input and methods with image-text input. As shown in Fig. \ref{visfig}, the first row shows that other methods occur in the upper part of the lung with different missing sizes, but our proposed C2FVL model can segment the lesion relatively completely. As seen in the second row, the results predicted by C2FVL are closer to the ground truth for the right lung region, which is more important for diagnosis.
As shown in Table \ref{tab1}, comparing the best-performing method nnUnet without text, C2FVL has a 3.51\% higher Dice score and 4.04\% higher IoU on the Qata-COVID dataset. Meanwhile, on the MosMedData+ dataset, the Dice score and IoU are improved by 1.97\% and 0.79\%, respectively. Compared with the baseline model with text like GLoRIA \cite{huang2021gloria}, C2FVL achieves all improvements of 3.46\%, 3.94\%, 2.14\% and 0.97\% under these two datasets in both two metrics.


\vspace{-4mm}
\subsection{Ablation Study}
\vspace{-2mm}
We perform a series of ablation experiments on the QaTa-COVID dataset for the associated hyper-parameters and different components of the model.

\begin{figure}[!ht]
\setlength{\abovecaptionskip}{-10cm}
\setlength{\belowcaptionskip}{-15cm}
\centering 
\includegraphics[width=0.95\linewidth]{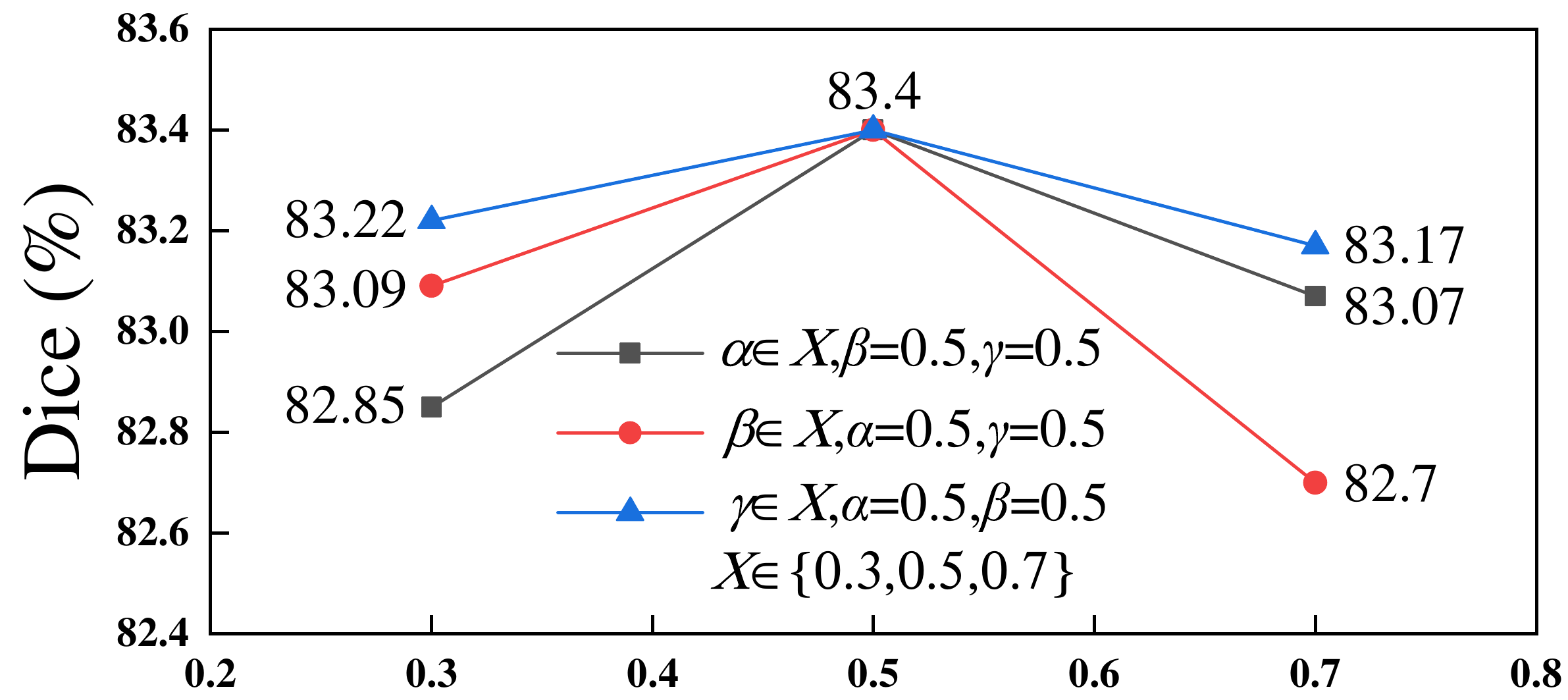}
\caption{Ablation study with different cosine loss coefficients. The gray, red and blue lines represent the trend of dice with $\alpha$, $\beta$ and $\gamma$ as variables, respectively.}
\label{Analysiscoef}
\vspace{-6mm}
\end{figure}

\noindent\textbf{Ablation study on different hyper-parameters:} For batch size, we selected three values of 2, 4 and 8 for comparison, and for learning rate, we set 1e-4, 1e-3, 1e-2. From the results in Table \ref{tab3}, we can see that when batch size is 4, learning rate is 1e-3, the model has the best Dice score and IoU value. For the loss coefficient,  we start with $\alpha=0.5$, $\beta=0.5$, $\gamma=0.5$, fixing two values among $\alpha$ , $\beta$, $\gamma$ as 0.5, and change the other value to observe trend of the model performance. As shown in Fig. \ref{Analysiscoef}, the optimal set of parameters is $\alpha=0.5$, $\beta=0.5$, $\gamma=0.5$.
\vspace{-2mm}
\vspace{-3mm}
\begin{table}[!ht]
\centering
\caption{Ablation study on different hyper-parameters.}
\resizebox{0.8\linewidth}{!}{%
\begin{tabular}{cccccc}
\hline
\multicolumn{1}{c}{\multirow{2}{*}{Hyper-Parameters}} & \multicolumn{3}{c}{\multirow{2}{*}{Value}}& \multicolumn{2}{c}{QaTa-COVID} \\ \cline{5-6} 
\multicolumn{2}{c}{}                    &            &  & Dice (\%)           & IoU (\%)               \\ \hline
\multirow{3}{*}{Batch Size}         &   \multicolumn{3}{c}{2}      &  82.03               &73.32             \\
                                       & \multicolumn{3}{c}{4}      & \textbf{83.40}         &\textbf{74.62}              \\
                                      &  \multicolumn{3}{c}{8}       & 83.32         & 74.57       \\ \hline
\multirow{3}{*}{Learning Rate}           &   \multicolumn{3}{c}{1e-4}                & 82.93 &74.19                     \\ &   \multicolumn{3}{c}{1e-3}           &\textbf{83.40} &\textbf{74.62} \\ 
& \multicolumn{3}{c}{1e-2}           &82.83 & 73.92 \\ \hline
\end{tabular}
}
\label{tab3}
\vspace{-8mm}
\end{table}

\begin{table}[!ht]
\centering
\caption{Ablation study on effectiveness of different components on QaTa-COVID dataset.}
\resizebox{\linewidth}{!}{%
\begin{tabular}{ccccccccc}
\hline
\multirow{2}{*}{Method} & \multirow{2}{*}{CNN} & \multicolumn{2}{c}{Text Input} & \multicolumn{2}{c}{C2FVLoss} & \multirow{2}{*}{VLAB} & \multirow{2}{*}{Dice (\%)} & \multirow{2}{*}{IoU (\%)} \\ \cline{3-6}
                        &                      & Single         & Multi         & $L_{V1}$            & $L_{V2}$           &                       &                     &                     \\ \hline
UNet++    & $\checkmark$                     &             &                          &            &              &              &                      79.62 &           70.25          \\ \hline
\multirow{4}{*}{C2FVL}  & $\checkmark$ &    $\checkmark$         &             &            &            &                        &  83.01                     & 74.26         \\
                        & $\checkmark$  &  &     $\checkmark$      &   &                 &            &  83.18   &   74.43                     \\
                        & $\checkmark$  & & $\checkmark$ & $\checkmark$ & & $\checkmark$  &  83.08 &  74.42                                    \\
& $\checkmark$& & $\checkmark$ & & $\checkmark$ & $\checkmark$  &  \textbf{83.40} &  \textbf{74.62}                    \\ \hline
\end{tabular}
}
\label{tab2}
\vspace{-2mm}
\end{table}

\noindent\textbf{Ablation study on effectiveness of different components: } As shown in Table \ref{tab2}, 
we use CNN that extracts image features as the backbone and then gradually added text information and VLAB. It is worth noting that when the single text is added to the original image input, the model's performance was substantially improved compared to the backbone, with Dice score improving by 3.39\% and IoU by 4.01\%. The model's performance is further improved after adding the corresponding scale text in each layer. On this basis, the performance of VLAB with cosine loss using the downsampling method decreases, but the Dice score and IoU values of VLAB with cosine loss using upsampling method reach the optimum. Therefore, we use $L_{V2}$ for the VLAB in the subsequent experiments.
\vspace{-4mm}

\begin{figure}[!ht]
	\centering
	\subfigure[GT]{
		\begin{minipage}[t]{0.18\linewidth}
			\centering
			\includegraphics[width=0.6in]{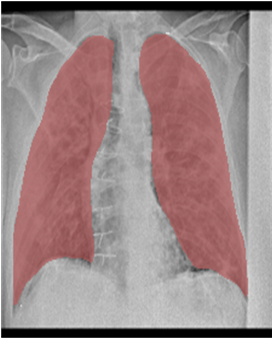}\\
			\vspace{0.01cm}
			\includegraphics[width=0.6in]{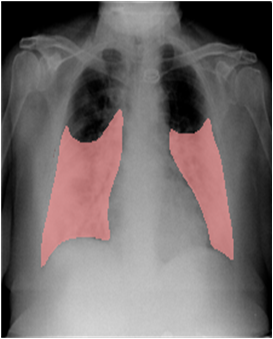}\\
			\vspace{0.01cm}
		\end{minipage}%
	}%
	\vspace{-3mm}
	\subfigure[CNN1]{
		\begin{minipage}[t]{0.18\linewidth}
			\centering
			\includegraphics[width=0.6in]{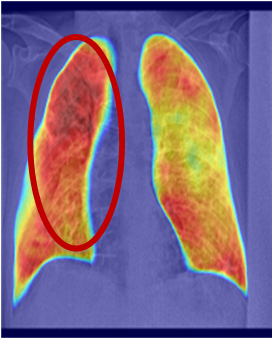}\\
			\vspace{0.01cm}
			\includegraphics[width=0.6in]{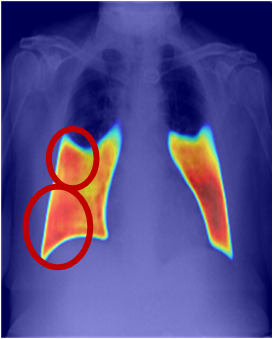}\\
			\vspace{0.01cm}
		\end{minipage}%
	}%
	\subfigure[CNN2]{
		\begin{minipage}[t]{0.18\linewidth}
			\centering
			\includegraphics[width=0.6in]{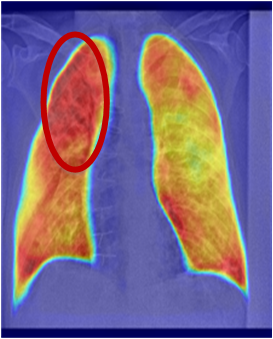}\\
			\vspace{0.01cm}
			\includegraphics[width=0.6in]{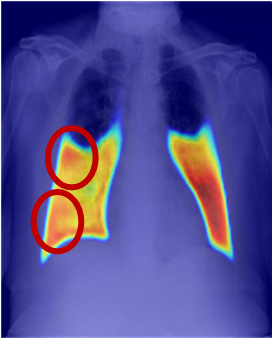}\\
			\vspace{0.01cm}
		\end{minipage}%
	}%
	\subfigure[CNN3]{
		\begin{minipage}[t]{0.18\linewidth}
			\centering
			\includegraphics[width=0.6in]{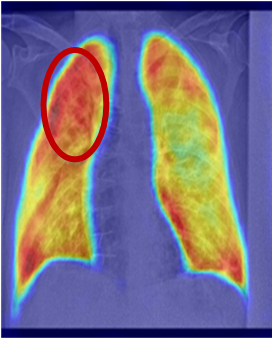}\\
			\vspace{0.01cm}
			\includegraphics[width=0.6in]{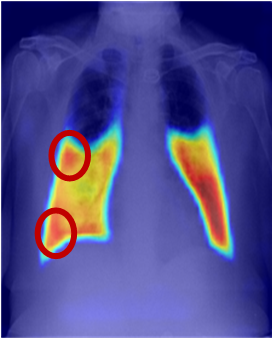}\\
			\vspace{0.01cm}
		\end{minipage}%
	}%
        \subfigure[CNN4]{
		\begin{minipage}[t]{0.18\linewidth}
			\centering
			\includegraphics[width=0.6in]{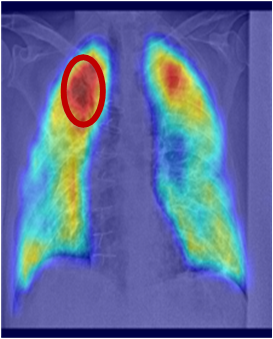}\\
			\vspace{0.01cm}
			\includegraphics[width=0.6in]{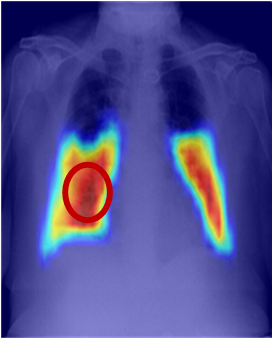}\\
			\vspace{0.01cm}
		\end{minipage}%
	}%
	\centering
	\caption{Saliency map for interpretability study on QaTa-COVID dataset. From (a) to (d) are the saliency maps corresponding to the outputs of four CNN decoders from bottom (CNN4) to top (CNN1) of different outputs in each CNN decoder. (e) ``GT'' represents the corresponding ground truth.}
	\vspace{-2mm}
	\label{vis}
\end{figure}
\vspace{-6mm}
\subsection{Interpretability Study}
\vspace{-3mm}
As shown in Fig. \ref{vis}, to verify that our cosine loss enhances the network's focus on the lesion region, we use Grad-CAM \cite{selvaraju2017grad} to show the activation region of the decoder part on the QaTa-COVID dataset. As seen from the left lung in the first row, the activation mapping of CNN4 shows a relatively narrow region of strong activation. As the layers of the network become progressively shallower, the region of strong activation begins to spread from the location of CNN4 to nearby regions, eventually focusing on the entire lesion region. For the left lung in the second row, the activation map of CNN4 shows the strong activation region mainly at the center of the lesion region, then splits to the edge of the left lung, followed by spreading from the edge to the central region. This result also demonstrates that the designed cosine loss in C2FVL can help the model detect the location of inaccurate focus and force it to refocus on the correct lesion region.
\vspace{-4mm}
\section{Conclusion}
\label{sec:conclusion}
\vspace{-3mm}
In this paper, we propose a new segmentation network called C2FVL, which uses text and image information in the training process. In addition, we design a VLAB to facilitate the alignment of image and text information and add cosine loss between different layers in the loss function to make each layer of the network focus on the lesion region. The experimental results show that our model improves performance compared to other state-of-the-art models. While our current work scratches the surface on multi-scale Vision Transformers for image classification, we anticipate that in future there will be more works in developing C2FVL for more applications, including more new released datasets.

\vspace{-2mm}
\section{Acknowledgement}
\vspace{-2mm}
This work was supported in part by the Natural Science Foundation of Fujian Province of China (No. 2020J01006), the Open Project Program of State Key Laboratory of Virtual Reality Technology and Systems, Beihang University (No.VRLAB2022AC04), and InnoHK. 


\bibliographystyle{IEEEbib}
\vspace{-2mm}
\bibliography{refs}

\end{document}